\documentclass[aps,prb,twocolumn,groupedaddress,showpacs]{revtex4-1}

\usepackage{amsmath}
\usepackage{amsfonts}
\usepackage{amssymb}
\usepackage{MnSymbol}
\usepackage{xspace}
\usepackage{wasysym}
\usepackage{graphicx}
\usepackage[utf8]{inputenc}
\usepackage[unicode]{hyperref}
\usepackage{color}
\usepackage{braket}
\usepackage{dsfont}

\setcitestyle{square,numbers}

\newcommand{\ie}[0]{i.e.\@\xspace}
\newcommand{\eg}[0]{e.g.\@\xspace}
\newcommand{\etal}[0]{\textit{et al.}\@\xspace}
\newcommand{\cf}[0]{cf.\@\xspace}

\newcommand{\fan}[1]{\hat{c}^{\vphantom\dagger}_{#1}}
\newcommand{\fcr}[1]{\hat{c}^{\dagger}_{#1}}
\newcommand{\fden}[1]{\hat{n}_{#1}}

\newcommand{\ban}[1]{\hat{b}^{\vphantom\dagger}_{#1}}
\newcommand{\bcr}[1]{\hat{b}^{\dagger}_{#1}}

\newcommand{\rhoan}[1]{\hat{\rho}^{\vphantom\dagger}_{#1}}
\newcommand{\rhocr}[1]{\hat{\rho}^{\dagger}_{#1}}

\newcommand{\Q}[1]{\hat{Q}_{#1}}
\renewcommand{\P}[1]{\hat{P}_{#1}}

\newcommand{\q}[1]{q_{#1}}
\newcommand{\p}[1]{p_{#1}}

\newcommand{\fcohan}[1]{c_{#1}}
\newcommand{\fcohcr}[1]{\bar{c}_{#1}}
\newcommand{\fcohmeasure}{\mathcal{D}(\bar{c},c)}

\newcommand{\bcohan}[1]{b_{#1}}
\newcommand{\bcohcr}[1]{\bar{b}_{#1}}
\newcommand{\bcohmeasure}{\mathcal{D}(\bar{b},b)}

\newcommand{\rhocohan}[1]{\rho_{#1}}
\newcommand{\rhocohcr}[1]{\bar{\rho}_{#1}}

\newcommand{\gencr}[1]{\bar{\eta}_{#1}}
\newcommand{\genan}[1]{\eta_{#1}}
\newcommand{\genP}[1]{\zeta_{#1}}
\newcommand{\genQ}[1]{\xi_{#1}}

\newcommand{\Pp}{P_{\! +}}
\newcommand{\Pm}{P_{\! -}}
\newcommand{\Ppm}{P_{\! \pm}}
\newcommand{\Pq}{P_{\! q}}
\newcommand{\Pnoq}{P}
\newcommand{\Ppmbar}{\bar{P}_{\!\pm}}

\newcommand{\omq}{\omega_q}
\newcommand{\omz}{\omega_0}

\newcommand{\Eekin}{E_{\mathrm{e\vphantom{ph}}}^{\mathrm{kin}\vphantom{\mathrm{pk}}}}
\newcommand{\Ephkin}{E_{\mathrm{ph}}^{\mathrm{kin}\vphantom{\mathrm{pk}}}}
\newcommand{\Ephpot}{E_{\mathrm{ph}}^{\mathrm{pot}\vphantom{\mathrm{pk}}}}
\newcommand{\Eeph}{E_{\mathrm{eph}}^{\vphantom{\mathrm{p}}}}
\newcommand{\Ephz}{E_{\mathrm{ph}}^{0}}
\newcommand{\Ephkintext}{E_{\mathrm{ph}}^{\mathrm{kin}}}
\newcommand{\Ephpottext}{E_{\mathrm{ph}}^{\mathrm{pot}}}
\newcommand{\Eephtext}{E_{\mathrm{eph}}}

\newcommand{\absolute}[1]{\left| #1 \right|}
\newcommand{\expv}[1]{\left\langle #1 \right\rangle}
\newcommand{\expvtext}[1]{\langle #1 \rangle}
\newcommand{\expvz}[1]{\left\langle #1 \right\rangle_0}
\newcommand{\expvztext}[1]{\langle #1 \rangle_0}
\newcommand{\expvc}[1]{\left\llangle #1 \right\rrangle_{C_n}}
\newcommand{\expvctext}[1]{\llangle #1 \rrangle_{C_n}}

\newcommand{\dtauobs}{\Delta\tau_{\mathrm{obs}}}

\begin{document}


\title{Continuous-time quantum Monte Carlo for fermion-boson lattice
  models: \\ 
  Improved bosonic estimators
and application to the Holstein model}


\author{Manuel Weber}
\author{Fakher F. Assaad}
\author{Martin Hohenadler}
\affiliation{\mbox{Institut f\"ur Theoretische Physik und Astrophysik,
Universit\"at W\"urzburg, 97074 W\"urzburg, Germany}}


\date{\today}

\begin{abstract}
We extend the continuous-time interaction-expansion quantum
Monte Carlo method with respect to measuring observables for
fermion-boson lattice models. Using generating functionals, we express expectation
values involving boson operators, which are not directly accessible because
simulations are done in terms of a purely fermionic action,
as integrals over fermionic correlation functions. We also demonstrate that
certain observables can be inferred directly from the vertex distribution, 
and present efficient estimators for the total energy and the
phonon propagator of the Holstein model. Furthermore, we generalize the
covariance estimator of the fidelity susceptibility, an unbiased diagnostic
for phase transitions, to the case of retarded interactions. 
The new estimators are applied to half-filled spinless and spinful Holstein
models in one dimension. The observed renormalization of the phonon mode
across the Peierls transition in the spinless model suggests a soft-mode
transition in the adiabatic regime. The critical point is associated with a
minimum in the phonon kinetic energy and a maximum in the fidelity susceptibility.
\end{abstract}


\maketitle

\section{Introduction}

Quantum Monte Carlo (QMC) methods are among the most established and
powerful tools to solve the quantum many-body problem of
correlated electrons. In particular, the auxiliary-field QMC method
\cite{PhysRevD.24.2278} and the stochastic series expansion 
(SSE) representation \cite{PhysRevB.43.5950} are widely used to
simulate lattice models, whereas more recent continuous-time (CTQMC)
methods \cite{PhysRevB.72.035122, PhysRevLett.97.076405} are
predominantly applied as impurity solvers in dynamical mean-field theory (DMFT)
\cite{RevModPhys.83.349}.   
Recently, progress has been made in the development of new
methods to simulate fermionic lattice models \cite{PhysRevB.91.241118,
  PhysRevB.91.235151}, the solution of the fermionic sign problem for specific
models \cite{PhysRevD.82.025007, Chandrasekharan2013,
  PhysRevB.91.241117, PhysRevLett.115.250601}, and the calculation of novel
observables such as the entanglement entropy
\cite{PhysRevLett.104.157201, PhysRevB.86.235116,
  PhysRevLett.111.130402, PhysRevB.89.125121, PhysRevB.91.125146,
  2014JSMTE..08..015B, PhysRevLett.113.110401,
  PhysRevB.92.125126}
and the fidelity susceptibility \cite{PhysRevE.76.022101,
  PhysRevLett.103.170501, PhysRevB.81.064418, PhysRevX.5.031007}.

For a large class of QMC methods (\eg, SSE and CTQMC), the partition function is calculated stochastically in a series
expansion and operators that are sampled can be measured directly from the Monte Carlo configurations.
In this paper, we consider the continuous-time interaction expansion
(CT-INT) method \cite{PhysRevB.72.035122}. In CT-INT, the
configurations are sets of interaction vertices and expectation
values are usually calculated from the single-particle Green's
function using Wick's theorem \cite{PhysRevB.81.024509}. However,
it can be advantageous to exploit the
information contained in the distribution of vertices, an important example
being the fidelity susceptibility \cite{PhysRevX.5.031007}.

The action-based formulation of the CT-INT method in particular allows
efficient simulations of fermion-boson lattice models \cite{PhysRevB.76.035116},
and has been successfully applied to electron-phonon problems
\cite{PhysRevB.83.115105,
  PhysRevLett.109.116407,PhysRevB.88.064303,PhysRevB.91.245147}. If the
action is quadratic in the bosonic fields, the latter can be
integrated out exactly \cite{PhysRev.97.660}, resulting in a fermionic action
with retarded interactions. Remarkably, autocorrelations, which
can be prohibitively strong in cases where the bosons are sampled
explicitly \cite{Hohenadler2008}, are significantly reduced in the fermionic representation.

An apparent disadvantage of the fermionic approach is the loss of access to bosonic observables.
However, as shown here, the latter can be systematically calculated from fermionic
correlation functions using sum rules derived from generating
functionals. Information about the bosonic fields is also 
encoded in the distribution of vertices. For a local
fermion-boson interaction (\eg, the Holstein model \cite{Ho59a}), the bosonic
contributions to the total energy as well as the local bosonic
propagator can be calculated efficiently from the vertex
distribution. Moreover, with the help of auxiliary Ising fields 
\cite{PhysRevB.28.4059} originally introduced to avoid the sign problem
\cite{PhysRevB.76.035116}, even nonlocal correlation functions such as
the full bosonic propagator become accessible.
Similar techniques have been applied to solve
fermion-boson problems with DMFT and the hybridization expansion
(CT-HYB) method \cite{PhysRevLett.97.076405} to understand 
dynamical screening effects \cite{PhysRevLett.104.146401,
  PhysRevB.89.235128, 2016arXiv160200584W}, and in extended DMFT
calculations \cite{PhysRevB.66.085120, PhysRevB.87.125149}. The usefulness of such techniques for computationally expensive lattice
problems was so far unclear but is demonstrated here. Finally, we derive an estimator for the 
fidelity susceptibility applicable to retarded boson-mediated
interactions that can be used to identify phase transitions.

We apply these (improved) estimators to one-dimensional Holstein
models \cite{Ho59a}. These fundamental models for the effects of
electron-phonon interaction constitute a significant
numerical challenge due to the infinite phonon Hilbert space, and the
different time scales for the fermion and boson dynamics.
In the half-filled case considered here, they describe a quantum phase
transition from a metallic phase to a Peierls insulator with long-range
charge-density-wave order \cite{PhysRevLett.49.402,PhysRevB.60.7950}.
We investigate two important open questions, namely, the renormalization of
the phonon spectrum across the Peierls transition in the adiabatic
regime, and two alternative diagnostics (phonon kinetic energy, fidelity
susceptibility) to locate the critical point. Importantly, our methodological
developments can also be applied in higher dimensions and for other models.

The paper is organized as follows. In Sec.~\ref{Sec:Method}, we discuss 
the calculation of observables from the vertex distribution in a general
formulation of the CT-INT method. In Sec.~\ref{Sec:PathInt}, we derive the
effective fermionic action for fermion-boson models and obtain estimators for
the total energy and the phonon propagator of the Holstein model. The calculation of bosonic observables from the vertex
distribution with the CT-INT method is discussed in
Sec.~\ref{Sec:CT-INT_conf}. A performance test and results for Holstein
models are presented in Sec.~\ref{Sec:Results}. We conclude in
Sec.~\ref{Sec:Conclusions}, and provide appendices on the relation between
bosonic observables and the dynamic charge-structure factor as well as on further improvements of the estimators.

\section{Quantum Monte Carlo method}
\label{Sec:Method}

\subsection{General formulation of the CT-INT method}

The CT-INT method \cite{PhysRevB.72.035122} is based on the
path-integral formulation of the grand-canonical
partition function
\begin{align}
\label{Eq:Z}
Z
  =
  \! \int \! \! \fcohmeasure \,
  e^{-S_0[\fcohcr{},\fcohan{}]-S_1[\fcohcr{},\fcohan{}]} \, ,
\end{align}
where the fermions are given in the Grassmann coherent-state representation
$\fan{}\ket{c} = \fcohan{}\ket{c}$ and time-ordering is implicit. We split the action into
the free-fermion part $S_0$ and the interaction $S_1$. The
weak-coupling perturbation expansion of Eq.~(\ref{Eq:Z}) is
\begin{align}
\label{Eq:Zexpansion_gen}
\frac{Z}{Z_0}
  =
  \sum_{n=0}^{\infty} \frac{\left(-1\right)^n}{n!} \expvz{S_1^n} \, ,
\end{align}
where we have defined $\expvztext{O} = Z_0^{-1} \int \fcohmeasure \,
e^{-S_0} O$ with $Z_0 = \int \fcohmeasure \, e^{-S_0}$. In the CT-INT
method, the expansion in Eq.~(\ref{Eq:Zexpansion_gen}) is
calculated stochastically by sampling configurations of interaction vertices. For this purpose, we
write the interaction in the vertex notation
\begin{align}
\label{Eq:Vertex_notation}
S_1
  =
    \sum_{\nu} w_{\nu}  h_{\nu} \, .
\end{align}
A vertex is represented by an instance of the superindex $\nu$ that
contains both discrete (\eg, lattice sites)
and continuous variables (\eg, imaginary times), a weight $w_{\nu}$, and the
Grassmann representation of the operators $h_{\nu}[\fcohcr{},\fcohan{}]$. 
The perturbation expansion becomes
\begin{align}
\label{Zexpansion}
\frac{Z}{Z_0}
  =
  \sum_{n=0}^{\infty}
  \underbrace{
  \sum_{\nu_1 \dots \nu_n}
   }_{\sum_{C_n}}
  \underbrace{
  \frac{\left(-1\right)^n}{n!} \,
  w_{\nu_1} \dots w_{\nu_n}
  \expvz{ h_{\nu_1} \dots h_{\nu_n}}
  \vphantom{  \sum_{n=0}^{\infty} \sum_{\nu_1 \dots \nu_n}}
  }_{W[C_n]} \, .
\end{align}
The sum runs over the expansion order $n$ and all
configurations of vertices $C_n = \{\nu_1,\dots,\nu_n\}$ for a given
$n$. We can identify the weight $W[C_n]$ to be sampled with the
Metropolis-Hastings algorithm
\cite{1953JChPh..21.1087M,10.2307/2334940}, which involves 
the determinant $\expvz{ h_{\nu_1} \dots h_{\nu_n}}=\det M[C_n]$ of
the $\mathcal{O}(n)\times \mathcal{O}(n)$ matrix $M[C_n]$ whose
entries are noninteracting Green's functions. Updates correspond
to the addition or removal of individual vertices, and involve
matrix-vector multiplications with $\mathcal{O}(n^2)$
operations. Since $\mathcal{O}(n)$ updates are necessary to reach an independent
configuration, the algorithm scales as $\mathcal{O}(n^3)$. The 
average expansion order $\langle n\rangle$ scales linearly with
the system size $L$ and the inverse temperature $\beta=(k_BT)^{-1}$
\cite{PhysRevB.72.035122} (see below). 
Expectation values $\expvtext{O} = Z^{-1} \int \fcohmeasure \, e^{-S_0-S_1} O$
are calculated via 
\begin{align}
\label{Eq:Obs_MC}
\expv{O} = \sum_{n=0}^{\infty} \sum_{C_n} \, p[C_n] \expvc{O} \, ,
\end{align}
where $p[C_n] = W[C_n]/ \sum_n \sum_{C_n} W[C_n]$
and $\expvctext{O}$ is the value of the observable for
configuration $C_n$. For any $C_n$, Wick's theorem
\cite{PhysRevB.81.024509} can be used to calculate $\expvctext{O}$ from the single-particle Green's
function. However, especially the calculation of the time-displaced
Green's function can be expensive because a matrix-vector
multiplication of $\mathcal{O}(n^2)$ must be performed for each
imaginary time $\tau$ and each pair of lattice sites. For further
details, see Ref.~\cite{RevModPhys.83.349}.

\subsection{Estimators from the vertex distribution}

In the SSE method \cite{PhysRevB.43.5950}, operators contained in
the operator string are accessible from the Monte Carlo configurations,
whereas in the CT-HYB method \cite{PhysRevLett.97.076405} the
single-particle Green's function can be obtained directly from the
perturbation expansion. Similarly, in CT-INT, expectation values of operators
$h_{\nu}$ contained in the interaction $S_1$ can be calculated
efficiently from the distribution of vertices
\cite{PhysRevB.56.14510}. To this end, $h_{\nu}$ is regarded as an
additional vertex written as $h_{\nu} = w_{\nu}^{-1}
\sum_{\nu_{n+1}} w_{\nu_{n+1}} h_{\nu_{n+1}} \delta_{\nu,\nu_{n+1}}$
and absorbed into the perturbation expansion:
\begin{align}
\expv{h_{\nu}}
  &=
    \frac{Z_0}{Z}
    \sum_{n=0}^{\infty}
\sum_{C_n}
    \frac{\left(-1\right)^n}{n!} \,
    w_{\nu_1} \dots w_{\nu_n}
    \expvz{ h_{\nu_1} \dots h_{\nu_n} h_{\nu} }
  \nonumber\\
  &=
    - \frac{1}{w_{\nu}}
    \sum_{n=0}^{\infty}
\sum_{C_{n+1}}
    \!\!
    \left( n+1 \right) p[C_{n+1}] \,
    \delta_{\nu,\nu_{n+1}}
\\
  &=
    \sum_{n=0}^{\infty}
    \sum_{C_n} \, p[C_n]
    \left[
    -\frac{1}{w_{\nu}}
    \sum_{k=1}^{n} \delta_{\nu,\nu_k}
    \right]
    \, .
    \nonumber
\end{align}
Here, we first identified the probability distribution $p[C_{n+1}]$ of a
configuration with $n+1$ vertices and then shifted the 
summation index to obtain $p[C_n]$. Finally, we included the $n=0$
contribution to the sum and replaced the factor of $n$ by a sum over
the equivalent vertices.
Comparison with Eq.~(\ref{Eq:Obs_MC}) yields 
\begin{align}
\label{Eq:Obs_vert_1}
\expvc{h_{\nu}} = - \frac{1}{w_{\nu}} \sum_{k=1}^{n} \delta_{\nu,\nu_k} \ .
\end{align}
From Eq.~(\ref{Eq:Obs_vert_1}) we obtain the familiar relation
between the interaction term and the average expansion order,
$\expv{S_1}=-\expv{n}$ \cite{PhysRevB.72.035122}. Because $\expv{S_1}$
is an extensive thermodynamic quantity, the average expansion order
$\expv{n} \sim \beta L$.
In the same way, we can obtain higher-order correlation functions, \eg,
\begin{align}
\label{Eq:Obs_vert_2}
\expvc{h_{\nu}h_{\nu'}} = \frac{1}{w_{\nu}w_{\nu'}} \sum_{k\neq l} \delta_{\nu,\nu_k} \delta_{\nu',\nu_l} \ .
\end{align}
Each variable contained in $\nu$ can be resolved from a configuration
$C_n$, but continuous variables (\eg, imaginary time $\tau$)
have to be integrated over (at least on a small interval) to make
sense of the corresponding delta functions.
The evaluation of observables via Eqs.~(\ref{Eq:Obs_vert_1}) and
(\ref{Eq:Obs_vert_2}) only requires $\mathcal{O}(n)$ operations since
\begin{align}
\label{Eq:sumtrick}
\sum_{k \neq l} f_{ik} f_{jl}
  =
  \sum_k f_{ik}
  \sum_l f_{jl}
  - \sum_k f_{ik} f_{jk} \, .
\end{align}
Because only operators that appear in the interaction can be measured,
the cheaper vertex measurements cannot completely
replace the more expensive calculation of the single-particle Green's
function.
However, the class of accessible observables grows with the
complexity of the interaction, as demonstrated below for
the fermion-boson problem.

\subsection{Fidelity susceptibility}

Recently, Wang \etal~\cite{PhysRevX.5.031007} derived a universal
QMC estimator for the fidelity susceptibility $\chi_\text{F}$ based on the
distribution of vertices. We briefly summarize their results, focusing on the
CT-INT method.

The fidelity susceptibility is a geometrical tool originating from
quantum information theory \cite{2008arXiv0811.3127G}. It can be used to
detect quantum critical points without prior knowledge of the order
parameter from the change of the ground state upon changing the
Hamiltonian $\hat{H}(\alpha) = \hat{H}_0 + \alpha \, \hat{H}_1$ via a
driving parameter $\alpha$. In Refs.~\cite{PhysRevE.76.022101,
  PhysRevLett.103.170501, PhysRevB.81.064418}, $\chi_\text{F}$ was
generalized to finite temperatures in terms of the structure factor
\begin{align}
\label{Eq:FS_finiteT}
\chi_\text{F}(\alpha)
  =
  \int_0^{\beta/2} \! \! d\tau
  \left[
          \expv{\hat{H}_1(\tau) \hat{H}_1(0)}
           - \expv{\hat{H}_1(0)}^2
  \right]
  \tau
\, .
\end{align}
Wang \etal~\cite{PhysRevX.5.031007} recognized that
Eq.~(\ref{Eq:FS_finiteT}) can be recovered from the distribution of
vertices using Eqs.~(\ref{Eq:Obs_vert_1}) and
(\ref{Eq:Obs_vert_2}), leading to the covariance estimator
\begin{align}
\label{Eq:FS_MC}
\chi_\text{F}
  =
  \frac{\expv{n_\text{L} n_\text{R}} - \expv{n_\text{L}} \expv{n_\text{R}}}{2\alpha^2} \, .
\end{align}
For each vertex configuration, ${n}_\text{L}$ and ${n}_\text{R}$ count the number of
vertices in the intervals $[0,\beta/2)$ and $[\beta/2,\beta)$, respectively. 
The calculation of $\chi_\text{F}$ via Eq.~(\ref{Eq:FS_MC}) is restricted to
fermionic actions that are local in time and related to a Hamiltonian,
\ie, $S_1 = \alpha \int \! d\tau H_1(\tau)$. A generalization to 
retarded boson-mediated interactions is given below.

\section{Path-integral formulation of the fermion-boson problem}
\label{Sec:PathInt}

In the following, we derive an effective fermionic action for a generic
fermion-boson model that can be simulated with the CT-INT method.
With the help of generating 
functionals, any bosonic observable can be recovered from fermionic
correlation functions. In particular, we derive sum rules for the
phonon propagator and the total energy of the Holstein model.

\subsection{Fermion-boson models}

We consider a generic one-dimensional fermion-boson Hamiltonian of the form
\begin{align}
\label{Eq:Ham_gen}
  \hat{H}
          =
             \hat{H}_0
             + \sum_q \omq  \bcr{q} \ban{q} 
             + \sum_q \gamma_q \left( \rhoan{q} \bcr{q} + \rhocr{q} \ban{q}   \right) \,,
\end{align}
with  fermionic (bosonic) creation and annihilation operators
$\fcr{}$, $\fan{}$ ($\bcr{}$, $\ban{}$) and the free-fermion
part $\hat{H}_0[\fcr{} \!\!, \fan{}]$. $\hat{H}$ is restricted to be quadratic in the bosons, but
we allow a general dispersion $\omega_q$ and a coupling to an arbitrary fermionic
operator $\rhoan{q}[\fcr{} \!\!, \fan{}]$ with coupling parameter $\gamma_q$. 

As an example, we consider the Holstein model
\cite{Ho59a}
\begin{align}
\label{Eq:Holstein_model}
\hat{H}
  =
\hat{H}_0
+\sum_i \left( \frac{1}{2M}  \P{i}^2 + \frac{K}{2}  \Q{i}^2 \right)
+ g \sum_i \Q{i} \hat{\rho}_i
\, ,
\end{align}
where the electronic part is given by the nearest-neighbor hopping
of spinful fermions with amplitude $t$,
\begin{align}
\hat{H}_0
  =
  -t \sum_{i\sigma} \left( \fcr{i\sigma} \fan{i+1\sigma}
                            + \fcr{i+1\sigma} \fan{i\sigma} \right) \, .
\end{align}
The phonons are described by local harmonic oscillators with
displacements $\Q{i}$ and momenta $\P{i}$; $M$ is the oscillator
mass and $K$ the spring constant. The displacements couple to the
charge density $\hat{\rho}_i= \sum_{\sigma} (
  \hat{n}_{i\sigma} - 1/2)$ (here
  $\fden{i\sigma}=\fcr{i\sigma}\fan{i\sigma}$) with coupling parameter
  $g$. The spinless Holstein model is obtained by dropping spin indices.

The Holstein model follows from the generic model~(\ref{Eq:Ham_gen}) by
dropping the momentum dependence of the bosons, \ie, $\omq\to\omz$ and $\gamma_q
\to\gamma$, and assuming a density-displacement coupling so that
$\rhocr{q} = \rhoan{-q}$. The same simplifications arise in
electron-phonon models with nonlocal density-displacement
\cite{PhysRevLett.109.116407} or bond-displacement couplings
\cite{PhysRevB.91.245147}. 
Therefore, the formulas derived below for the Holstein model
can be easily transferred to other models. For the Holstein case, $\omz
= \sqrt{K/M}$, $\gamma = g/\sqrt{2M\omz}$, and we also 
introduce the dimensionless coupling parameter $\lambda =
\gamma^2/(2\omega_0t) = g^2/(4Kt)$. Simulations were
performed at half-filling, but the estimators are general.

\subsection{Effective fermionic action for the bosons and observables
  from generating functionals}

For the generic fermion-boson model (\ref{Eq:Ham_gen}), the partition
function takes the form 
\begin{align}
\label{Eq:pathint_eph}
  Z
    = \!
      \int \! \! \fcohmeasure \, e^{-S_0[\fcohcr{},\fcohan{}]} \!
      \int \! \! \bcohmeasure \,
  e^{-S_{\mathrm{ep}}[\fcohcr{},\fcohan{},\bcohcr{},\bcohan{}]} \, .
\end{align}
We use the coherent-state representation $\fan{} \ket{c} =
\fcohan{} \ket{c}$ with Grassmann variables $\fcohan{}$ for the
fermions, and $\ban{} \ket{b} = \bcohan{} \ket{b}$ with complex
variables $\bcohan{}$ for the bosons.
The action is split into the fermionic part $S_0$ and the remainder
$S_{\mathrm{ep}}$ containing the free-boson part and the
interaction,
\begin{align}
\label{Eq:Sep}
  S_{\mathrm{ep}}
    = \!
      \int_0^{\beta} \! \! d\tau \sum_q
      &\left\{ \,
	 \bcohcr{q}(\tau)
         \left[ \partial_{\tau} + \omq\right]
         \bcohan{q}(\tau)
         \right. \\  
	& \, \, \left.
          + \gamma_q
          \left[ \rhocohan{q}(\tau) \, \bcohcr{q}(\tau) +
          \rhocohcr{q}(\tau) \, \bcohan{q}(\tau) \right]
      \right\} \, .
\nonumber
\end{align}
The bosons can be integrated out exactly \cite{PhysRev.97.660},
leading to an effective fermionic interaction
\begin{align}
\label{Eq:Seff_gen}
  S_1
    =
      - \sum_q \frac{\gamma_q^2}{\omq}
         \iint_{0}^{\beta} d\tau d\tau'
	\rhocohcr{q}(\tau) \,
          \Pq(\tau-\tau') \, \rhocohan{q}(\tau')
\end{align}
mediated by the noninteracting bosonic Green's function
$\Pq(\tau-\tau') = \omq \expvztext{\bcohcr{q}(\tau)\bcohan{q}(\tau')}$.
Here, $\expvztext{\dots}$ also denotes expectation values
with respect to the free-boson part of the action. For  $0 \leq \tau < \beta$,
$\Pq(\tau)$ is given by
\begin{align}
\label{Eq:ph_green_q}
  \Pq(\tau)
    =
      \omq \,\frac{ e^{-\omq \tau}}{1-e^{-\omq \beta}}       
\end{align}
and we impose $\Pq(\tau+\beta)=\Pq(\tau)$.
With the factor of $\omq$, the adiabatic and antiadiabatic
limits of $\Pq(\tau)$ are
\begin{align}
\label{Eq:ph_green_limits}
  \lim_{\omq\to0} \Pq(\tau) = \frac{1}{\beta} \, , \qquad
  \lim_{\omq\to\infty} \Pq(\tau) = \delta(\tau) \, .
\end{align}

In principle, the fermionic interaction (\ref{Eq:Seff_gen}) can be
simulated with the CT-INT method if transformed into real
space. However, for any nontrivial dispersion $\omq$ the transformed
bosonic propagator has negative contributions that cause a sign
problem \cite{PhysRevB.91.245147}. Therefore, we focus on models with optical
bosons, \ie,  $\omq=\omz$.

To obtain estimators for bosonic correlation functions in the CT-INT method, 
we add the source term
\begin{align}
\label{Eq:gen_func_gen}
  S_{\mathrm{source}}
    =
      - \! \int_0^{\beta} \! \! d\tau \sum_q
	\left[ \genan{q}(\tau) \, \bcohcr{q}(\tau) + \gencr{q}(\tau) \, \bcohan{q}(\tau) \right]
\end{align}
to $S_{\mathrm{ep}}$.
After integrating out the bosons, the complex source
fields $\genan{q}(\tau)$ and $\gencr{q}(\tau)$ appear in $S_1$, \ie,
\begin{align}
\label{Eq:Seff_gen_func}
  S_{1,\mathrm{source}}
    =
      - \sum_q \frac{\gamma_q^2}{\omq} \iint_0^{\beta} &d\tau d\tau' 
        \left[\rhocohcr{q}(\tau) - \gamma_q^{-1} \gencr{q}(\tau) \right] \ \phantom{.} \\
	\times \, &\Pq(\tau-\tau')
	\left[\rhocohan{q}(\tau') - \gamma_q^{-1} \genan{q}(\tau')
  \right] \, .
\nonumber
\end{align}
From Eq.~(\ref{Eq:Seff_gen_func}), any bosonic correlation function
can be expressed in terms of fermionic fields by taking functional
derivatives and the limit $\eta\to0$.

\subsection{Application to the Holstein model}\label{sec:obs:hol}

In the following, we illustrate the use of this formalism for the
Holstein model (\ref{Eq:Holstein_model}). The notation is
kept as general as possible to facilitate applications to other models.
Replacing $\Pq(\tau)\to \Pnoq(\tau)$ the effective interaction
\begin{align}
\label{Eq:Seff_HS}
  S_1
    =
      - 2\lambda t
         \iint_0^{\beta} d\tau d\tau' \sum_i
	\rhocohan{i}(\tau)
         \, \Pnoq(\tau-\tau') \, \rhocohan{i}(\tau')
\end{align}
becomes diagonal in real space.
To express bosonic observables in terms of the displacements
$\q{i}(\tau)$ or the momenta $\p{i}(\tau)$
we rewrite the source term (\ref{Eq:gen_func_gen}) as
\begin{align}
  S_{\mathrm{source}}
    =
      - \! \int_0^{\beta} \! \! d\tau \sum_i
      \left[
	\genQ{i}(\tau) \, \q{i}(\tau) + \genP{i}(\tau) \, \p{i}(\tau)
       \right] \, ,
\label{gen1}
\end{align}
with real fields $\genQ{i}(\tau)$ and $\genP{i}(\tau)$.
Transformation of the source fields in Eq.~(\ref{Eq:Seff_gen_func})
leads to the action
\begin{align}
  S_{1,\mathrm{source}}
    =
      S_1
      + S_{\xi\rho}^+
      + S_{\xi\xi}^+
      + S_{\zeta\rho}^-
      + S_{\zeta\zeta}^+
      + S_{\xi\zeta}^- \, ,
\label{Eq:genac2}
\end{align}
where the individual contributions are given by
\begin{align}
\label{Eq:gen_precise}
S_{\mu\nu}^{\pm}
  =
  - \alpha_{\mu\nu}
  \iint_0^{\beta} d\tau d\tau' \sum_i
  \mu_i(\tau) \, \Ppm(\tau-\tau') \, \nu_i(\tau')
\end{align}
with
$\alpha_{\xi\rho} = -2 \sqrt{\lambda t/K}$,
$\alpha_{\xi\xi} = 1/(2K)$,
$\alpha_{\zeta\rho} = 2i \sqrt{M\lambda t}$,
$\alpha_{\zeta\zeta} = M/2$, and
$\alpha_{\xi\zeta} = i/\omz$.
Here, we defined
the phonon propagators
$\Ppm(\tau) = \frac{1}{2}\left[ \Pnoq(\tau) \pm \Pnoq(\beta-\tau) \right]$,
corresponding to
$\Pp(\tau-\tau') = K \, \expvztext{\q{i}(\tau) \q{i}(\tau')} = M^{-1}
\expvztext{\p{i}(\tau) \p{i}(\tau')}$ and
$\Pm(\tau-\tau') = -i \, \omz \expvztext{\q{i}(\tau) \p{i}(\tau')}$.

With the help of the generating functionals in Eqs.~(\ref{Eq:genac2})
and (\ref{Eq:gen_precise}), we get access to the phonon propagators
\begin{align}
\label{Eq:ph_prop_int_Q}
K \expv{\q{i}(\tau)\q{j}(\tau')}
  &=
        \Pp(\tau-\tau') \, \delta_{i,j} 
        +  X_{ij}^{++}(\tau,\tau') \, ,
\\
\label{Eq:ph_prop_int_P}
\frac{1}{M} 
\expv{\p{i}(\tau)\p{j}(\tau')}
  &=
        \Pp(\tau-\tau') \, \delta_{i,j} 
        +  X_{ij}^{--}(\tau,\tau') 
\end{align}
consisting of the free propagator $\Pp$ and the interaction contributions
\begin{align}
X_{ij}^{\pm\pm}(\tau,\tau')
=
  4\lambda t
  \iint_0^{\beta} d&\tau_1 d\tau'_1 \, 
        \Ppm(\tau-\tau_1) \\
       &\times  \expv{\rhocohan{i}(\tau_1) \rhocohan{j}(\tau'_1)}
        \Ppm(\tau'_1 - \tau') \, .
\nonumber
\end{align}
The total energy is $E = \Eekin + \Ephkin + \Ephpot + \Eeph$,
with
{\allowdisplaybreaks
\begin{align}
\Ephkin
    &=
      \frac{\Ephz}{2}
      - 2 \lambda t \iint_0^{\beta} d\tau d\tau'
      \Pm(\tau) \, \Pm(\tau') \, C_{\rho}(\tau-\tau') \, ,
	      \label{Eq:Epkin}
	\\
\Ephpot
    &=
      \frac{\Ephz}{2}
      + 2 \lambda t \iint_0^{\beta} d\tau d\tau'
      \Pp(\tau) \, \Pp(\tau') \, C_{\rho}(\tau-\tau') \, ,
	      \label{Eq:Eppot}
	\\
\Eeph
    &=
      - 4 \lambda t \! \int_0^{\beta} \! \! d\tau \,
      \Pp(\tau) \, C_{\rho}(\tau) \, .
	      \label{Eq:Eep}
\end{align}
}
Here, $\Ephz = L \Pp(0)$ and $C_{\rho}(\tau-\tau') = \sum_i
\expv{\rhocohan{i}(\tau)\rhocohan{i}(\tau')}$.
$\Ephpottext$ and $\Ephkintext$  follow from
Eqs.~(\ref{Eq:ph_prop_int_Q}) and (\ref{Eq:ph_prop_int_P}) by fixing
the interaction to $X^{\pm\pm}_{ii}(0,0)$. In Appendix
\ref{Sec:SumRulesEnergies}, we provide further information on the
relation between the bosonic observables and the dynamic charge
structure factor. 

The observables~(\ref{Eq:ph_prop_int_Q})--(\ref{Eq:Eep}) can be recovered
from the charge correlation function $\langle\rho_i(\tau)\rho_j(\tau')\rangle$
which is accessible in CT-INT via Wick's theorem. In Ref.~\cite{PhysRevB.91.235150},
we calculated $\expv{\rhocohan{i}(\tau)\rhocohan{j}(0)}$ on an
equidistant $\tau$ grid with spacing $\dtauobs=0.1$ and
performed the remaining integrals numerically. However, as shown
below, it is more efficient to use the distribution of vertices.

\section{CT-INT for the Holstein model}
\label{Sec:CT-INT_conf}

\subsection{Vertex notation for the effective interaction}

For the Holstein model, the interaction term sampled with the CT-INT method
takes the form
\begin{align}
  S_1
    =
      - \lambda t \iint_0^{\beta} d\tau d\tau' \!\! \sum_{i\sigma\sigma's}
				&\left[\rhocohan{i\sigma}(\tau) - s \delta \right] \\
				 &\times \Pp(\tau-\tau')
                                 \left[\rhocohan{i\sigma'}(\tau') - s
                                  \delta \right]
                                \, .
\nonumber
\end{align}
Compared to Eq.~(\ref{Eq:Seff_HS}), we introduced an auxiliary Ising
variable $s=\pm 1$ (and $\delta = 0.51$) to avoid the
sign problem \cite{PhysRevB.76.035116}, and used the symmetrized phonon
propagator $\Pp(\tau)$. In the notation of
Eq.~(\ref{Eq:Vertex_notation}),
$\nu=\{i,\tau,\tau',\sigma,\sigma',s\}$, $w_{\nu} = -\lambda t \,
\Pp(\tau-\tau')$, and 
\begin{align}
\label{Eq:HS_vert_1}
h_{\nu} = \rhocohan{i\sigma}(\tau) \rhocohan{i\sigma'}(\tau')
               + \delta^2 
               - s \delta \left[ \rhocohan{i\sigma}(\tau)
                                + \rhocohan{i\sigma'}(\tau') \right]
  \, .
\end{align}
The QMC simulation is performed
as described before. The acceptance rate for adding a new vertex can
be optimized by proposing $\tau-\tau'$ according
to $\Pp(\tau-\tau')$ via inverse transform sampling.

\subsection{Observables from the distribution of vertices}

The operators contained in Eq.~(\ref{Eq:HS_vert_1}) can be
measured from the distribution of vertices. In particular, we have
access to the dynamical charge correlations required for the calculation
of the bosonic observables in Sec.~\ref{sec:obs:hol}. In the following, we use
Eqs.~(\ref{Eq:Obs_vert_1}) and~(\ref{Eq:Obs_vert_2}) to derive
improved estimators for the total energy, the fidelity susceptibility,
and the phonon propagator.

\subsubsection{Total energy}

The kinetic energy of the electrons is calculated from the
single-particle Green's function. To recover the phononic
contributions (\ref{Eq:Epkin})--(\ref{Eq:Eep}) from the distribution
of vertices, we sum over the auxiliary Ising variable $s$ in
Eq.~(\ref{Eq:HS_vert_1}) and use Eq.~(\ref{Eq:Obs_vert_1}) to obtain
the estimator
\begin{align}
\label{Eq:den_loc_vert}
&\expvc{\rhocohan{i\sigma}(\tau) \rhocohan{i\sigma'}(\tau')}
+ \delta^2
\\
&\hspace{1.4cm}
=
 \sum_{k=1}^n
\frac{\delta_{i,i_k} \delta_{\sigma,\sigma_k}
  \delta_{\sigma' \!\!,\sigma'_k} \delta(\tau-\tau_k) \,
  \delta(\tau'-\tau'_k)}{2\lambda t \, \Pp(\tau_k -\tau'_k)} 
\nonumber
\end{align}
for the local charge-charge correlation function. From
Eq.~(\ref{Eq:den_loc_vert}) we get the estimators
\begin{align}
  \Ephkin[C_n]
    &=
      \frac{\Ephz}{2}
      - \sum_{k=1}^{n}
      \frac{\Pm(\tau_k) \Pm(\tau'_k)}{\Pp(\tau_k - \tau'_k)} \, ,
      \label{Eq:Epkin_vert}
	\\
  \Ephpot[C_n]
    &=
      \frac{\Ephz}{2}
      + \sum_{k=1}^{n}
      \frac{\Pp(\tau_k) \Pp(\tau'_k)}{\Pp(\tau_k - \tau'_k)}
      - {2\lambda t L N_{\sigma}^2\delta^2}     \, ,
      \label{Eq:Eppot_vert}
	\\
  \Eeph[C_n]
    &=
      - \frac{2 n}{\beta} + 4 \lambda t L N_{\sigma}^2\delta^2 \, .
            \label{Eq:Eep_vert}
\end{align}
For the kinetic energy the term $\sim\delta^2$ vanishes due to the
antisymmetry of $\Pm(\tau)$. $N_{\sigma}$ counts the number of spin
components of the Holstein model, \ie, $N_{\sigma}=1$ for the spinless
and $N_{\sigma}=2$ for the spinful model.

The estimators~(\ref{Eq:Epkin_vert}) and (\ref{Eq:Eppot_vert}) can
be further improved by exploiting the global translational invariance of all
vertices, \ie,
$\tau_k\to \tau_k+\Delta\tau$ and $\tau'_k\to \tau'_k+\Delta\tau$ with
$\Delta\tau\in[0,\beta)$. We integrate over $\Delta\tau$ to treat all
the translations exactly, see Appendix~\ref{Sec:TranslVert} for details.
Thereby, especially $\Ephkintext[C_n]$ is substantially improved, as
shown in Sec.~\ref{Sec:PerformanceTest}.

\subsubsection{Fidelity susceptibility}

To calculate the fidelity susceptibility for a retarded interaction
we start from Eq.~(\ref{Eq:FS_finiteT}) and identify the
electron-phonon coupling as the driving term with $\alpha=g$ and $\hat{H}_1 =
\sum_i \Q{i} \hat{\rho}_i$. The displacements $\Q{i}$ entering the
expectation values of the Hamiltonian in Eq.~(\ref{Eq:FS_finiteT}) can
be replaced with fermionic operators using the source terms introduced
before. $\expvtext{H_1}$ is given by Eq.~(\ref{Eq:Eep}), and
\begin{align}
\begin{split}
\expv{H_1(\tau) H_1(\tau')}
  =
  2 \sum_{\nu_1} w_{\nu_1} \expv{h_{\nu_1}} \delta(\tau-\tau_1) \,
  \delta(\tau'-\tau'_1) \\
+ 4 \sum_{\nu_1 \nu_2} w_{\nu_1} w_{\nu_2} \expv{h_{\nu_1} h_{\nu_2}}
  \delta(\tau - \tau'_1) \, \delta(\tau' - \tau'_2)
\end{split}
\end{align}
in the vertex notation of the Holstein model.
Continuing the derivation as in
Ref.~\cite{PhysRevX.5.031007}, we obtain an estimator very similar to
Eq.~(\ref{Eq:FS_MC}),
\begin{align}
\label{Eq:FS_MC_HS}
\chi_\text{F}
  =
  \frac{
          \expv{\tilde{n}_\text{L} \tilde{n}_\text{R}}
           - \expv{\tilde{n}_\text{L}} \expv{\tilde{n}_\text{R}}
  }{2g^2}
\, .
\end{align}
However, in the present case, each vertex contains two bilinears with
times $\tau_k$ and $\tau'_k$, and $\tilde{n}_\text{L}$ and $\tilde{n}_\text{R}$ count the numbers
of these bilinears in the left and right half of the partitioned
imaginary-time axis. For simplicity, we omitted a constant shift in
Eq.~(\ref{Eq:FS_MC_HS}) that arises from the $\delta$-dependent terms in
Eq.~(\ref{Eq:HS_vert_1}). Taking it into account leads to 
$\chi_\text{F} \to \chi_\text{F} - {2\lambda t L N_{\sigma}^2 \delta^2
\tanh(\beta\omz/4)}/{(\omz g^2)}$.

\subsubsection{Phonon propagator}

Equation (\ref{Eq:den_loc_vert}) only gives access to local
charge-charge correlations.
For the Holstein model, we can also obtain nonlocal correlation
functions from the distribution of vertices, including the phonon
propagator. For this purpose, we
exploit the information provided by the Ising variable $s$. If we
consider $\sum_s s \, h_{\nu}$, the first two terms in
Eq.~(\ref{Eq:HS_vert_1}) drop out and only individual charge
operators are left. Analogously, by taking
\begin{align}
\label{Eq:nonloc_vert}
\begin{split}
 \sum_{s_1 s_2} s_1 s_2 \, h_{\nu_1} h_{\nu_2}
=
4 \delta^2
&\left[ \rhocohan{i_1\sigma_1}(\tau_1) + \rhocohan{i_1\sigma'_1}(\tau'_1) \right] \\
\times &\left[ \rhocohan{i_2\sigma_2}(\tau_2) + \rhocohan{i_2\sigma'_2}(\tau'_2)  \right]
\, ,
\end{split}
\end{align}
we can recover nonlocal charge correlations from
Eq.~(\ref{Eq:Obs_vert_2}). The simplest estimator is the
charge susceptibility
\begin{align}
\label{Eq:suscharge_vert}
\chi_{ij}[C_n]
  &=
  \frac{1}{\beta}
  \iint d\tau d\tau'
  \expvc{\rhocohan{i}(\tau)\rhocohan{j}(\tau')}
  \\
  &= 
  \frac{1}{16 (\lambda t)^2 N_{\sigma}^2 \delta^2 \beta^3} 
  \sum_{k \neq l} 
                       \frac{s_k \, \delta_{i,i_k}}{\Pp(\tau_k-\tau'_k)}
                        \frac{s_l \, \delta_{j,i_l}}{\Pp(\tau_l-\tau'_l)}
    \, ,
    \nonumber
\end{align}
which is obtained from the summation over all variables except for the
lattice sites. Similarly, the (spin-resolved) charge
correlation function can be calculated directly in Matsubara
frequencies. The phonon propagators (\ref{Eq:ph_prop_int_Q}) and
(\ref{Eq:ph_prop_int_P}) take the form
\begin{widetext}
\begin{align}
\label{Eq:ph_prop_Q_vert}
K \expvc{q_i(\tau) q_j(\tau')}
  &=
  \Pp(\tau-\tau') \, \delta_{i,j}
  + \frac{1}{4\lambda t N_{\sigma}^2 \delta^2} 
  \sum_{k \neq l}
  \frac{\Pp(\tau-\tau_k) \Pp(\tau-\tau'_k) \, s_k \, \delta_{i,i_k}}{\Pp(\tau_k - \tau'_k)}
  \frac{\Pp(\tau'-\tau_l) \Pp(\tau'-\tau'_l) \, s_l \, \delta_{j,i_l}}{\Pp(\tau_l - \tau'_l)}
  \, ,
\\
\label{Eq:ph_prop_P_vert}
\frac{1}{M} \expvc{p_i(\tau) p_j(\tau')}
  &=
  \Pp(\tau-\tau') \, \delta_{i,j}
  - \frac{1}{\lambda t N_{\sigma}^2 \delta^2 \beta^2} 
  \sum_{k \neq l}
  \frac{\Pm(\tau-\tau_k) \, s_k \, \delta_{i,i_k}}{\Pp(\tau_k - \tau'_k)}
  \frac{\Pm(\tau'-\tau_l) \, s_l \, \delta_{j,i_l}}{\Pp(\tau_l - \tau'_l)} 
    \, .
\end{align}
\end{widetext}
To arrive at Eq.~(\ref{Eq:ph_prop_Q_vert}), we multiplied Eq.~(\ref{Eq:nonloc_vert})
with the symmetrized propagator $\Pp$ for each of the four times on
the right-hand side before integrating over the imaginary times.
For Eq.~(\ref{Eq:ph_prop_P_vert}), we included the
antisymmetrized propagator $\Pm$ only for one pair of times, but the
estimator can be further improved by considering the remaining three
combinations. Similar to $\Ephkintext[C_n]$, the estimator
(\ref{Eq:ph_prop_P_vert}) can be substantially improved by exploiting
translational invariance of the vertices, see Appendix~\ref{Sec:TranslVert}.

\section{Results}
\label{Sec:Results}

\subsection{Performance of the vertex measurements}
\label{Sec:PerformanceTest}

In the CT-INT method, the computation of the single-particle Green's function
for the calculation of observables via Wick's theorem requires
$\mathcal{O}(n^2LN_{\tau})$ operations, where $N_{\tau}$ is the number 
of $\tau$ points. If $N_\tau$ is scaled with $\beta$, the calculation
of dynamical correlation functions
is of the same order as the Monte Carlo updates.
For fermion-boson problems, even the bosonic energies in
Eqs.~(\ref{Eq:Epkin})--(\ref{Eq:Eep}) require the full time dependence of
$\expv{\rhocohan{i}(\tau) \rhocohan{j}(0)}$. On the other hand, the
calculation from the vertex distribution involves only $\mathcal{O}(n)$
operations for the energies and $\mathcal{O}(n N_{\tau})$ for the phonon
propagator. For the latter, exploiting translational invariance leads to
another $\mathcal{O}(L^2N_{\tau}^2)$ operations to set up the final
estimator, \cf Appendix \ref{Sec:TranslVert}. For large $n$, the computational
cost for the vertex measurements becomes negligible.

The above considerations were verified for the spinless Holstein model with
$\omz/t=0.4$, $L=\beta t = 22$, $\lambda=1.5$, and $1000$ QMC steps between
measurements. The average expansion order was
$\expv{n} \approx 660$ and we used $\dtauobs=0.1$ ($N_{\tau} =220$). The computation of dynamical correlation functions using Wick's
theorem took $26\%$ of the total time, of which $86\%$ went into the
matrix-vector multiplications necessary to calculate the Green's
function. Only $1\%$ of the total time was used for the
vertex measurements, most of which went into the
$\mathcal{O}(L^2N_{\tau}^2)$ operations necessary to set up the
translation-invariant phonon propagator. If we omitted this last
operation, the vertex measurements only took $0.02\%$ of the total
time, and were dominated by the exact evaluation of $\Ppm(\tau)$
for each vertex. Approximately the same time would be needed for
equal-time measurements from Wick's theorem using $N_{\tau}=1$. Hence,
further improvements through tabulation of $\Ppm(\tau)$ seem unnecessary.

Aside from the significant speed-up, another advantage of
the vertex measurements is the exact calculation of imaginary-time
integrals. In contrast, Wick's theorem provides
$\expv{\rhocohan{i}(\tau) \rhocohan{j}(0)}$ only on a finite grid so
that systematic errors from numerical integration can arise. For $\omz/t=0.4$, using Simpson's rule on an equidistant
grid with $\dtauobs=0.1$ was sufficient to make systematic errors irrelevant.
However,  more elaborate integration schemes may be necessary for larger $\omz$.

Table~\ref{Tab:Statistics} reports ratios of statistical errors of
averages obtained from either the vertex distribution or Wick's theorem,
as determined in the same simulation and hence for the same number of bins.
We considered different bosonic energies, as well as the charge
susceptibility $\chi(q)$ at $q=\pi$ which tracks charge-density-wave order.
For $\Ephpottext$ and $\Ephkintext$ we compared three different estimators:
the simple estimators~(\ref{Eq:Epkin_vert}) and~(\ref{Eq:Eppot_vert}) from one set
of vertices, the improved estimators using translational invariance
[Eq.~(\ref{Eq:prop_repl})], and the estimators for the phonon
propagators using the Ising spins, Eqs.~(\ref{Eq:ph_prop_Q_vert}) and~(\ref{Eq:ph_prop_P_vert}).

The reference results are for the spinless Holstein model with
$\omz/t=0.4$, $\lambda = 0.5$,  $L=\beta t = 22$, and $\delta =
0.51$. For the resulting rather small expansion order $\expv{n} \approx 151$,
the estimators from Wick's theorem have better statistics, \ie, the ratios in
Table~\ref{Tab:Statistics} are larger than one. The vertex estimators
improve significantly upon exploiting translational invariance, especially
$\Ephkintext$. Increasing the number of vertices per phase-space
volume via the interaction parameter $\lambda$ levels out the
differences between estimators, except for $\Ephkintext$ at
$\lambda=1.5$. In contrast, changing $\expv{n}$ via the
phase-space parameters $L$ and $\beta$ leaves most of the ratios essentially unchanged. The same is true when increasing the
number of vertices via the Ising-spin parameter $\delta$.
Finally, Table~\ref{Tab:Statistics} confirms that $\expv{n}\sim \beta L$,
whereas the dependence on $\lambda$ is nonlinear.
 \begin{table}
 \caption{\label{Tab:Statistics}
Ratios of statistical errors for averages from vertex measurements and
Wick's theorem for different simulation parameters. The reference point is
the spinless Holstein model with $\omz/t=0.4$, $\lambda = 0.5$,  $L=\beta t =
22$, and $\delta = 0.51$. The first two rows indicate the observable and 
estimator used. The last column reports the average expansion order.
}
 \begin{ruledtabular}
 \begin{tabular}{c||c|ccc|ccc|c||c}
observable   & $\Eephtext$ & \multicolumn{3}{c|}{$\Ephpottext$} 
         &\multicolumn{3}{c|}{$\Ephkintext$} & $\chi(\pi)$ & $\expv{n}$ \\
 from Eq.              & (\ref{Eq:Eep_vert}) &  (\ref{Eq:Eppot_vert}) &
            (\ref{Eq:prop_repl}) & (\ref{Eq:ph_prop_Q_vert}) & (\ref{Eq:Epkin_vert}) &
           (\ref{Eq:prop_repl}) & (\ref{Eq:ph_prop_P_vert})  &
           (\ref{Eq:suscharge_vert}) &  \\
\hline
    reference                        & 2.6 & 4.0 & 2.6 & 2.5 & 20 & 5.6 & 5.9 & 1.2 & 151\\
\hline
   $\lambda=1.0$ & 1.2 & 1.4 & 1.1 & 1.2 & 4.8 & 1.6 & 1.3 & 1.0 & 371 \\
   $\lambda=1.5$ & 1.1 & 1.3 & 1.1 & 1.6 & 18  & 3.3 & 2.9 & 1.0 & 661 \\
\hline 
   $L = \beta t =14$ & 3.2 & 4.0 & 3.2 & 0.2 & 19 & 6.4 & 4.0  & 1.2 & 62\\
   $L = \beta t =30$ & 2.6 & 5.0 & 2.7 & 2.8 & 23 & 5.4 & 13 & 1.3 & 282\\
\hline
   $\delta=1.0$ & 3.7  & 7.0 &  4.0 & 2.1 & 32 & 8.5 & 4.4 & 1.2 & 510\\
 \end{tabular}
 \end{ruledtabular}
 \end{table}

Although the dependence of the statistical errors on the simulation
parameters is not completely systematic, the vertex measurements
become advantageous especially at large expansion orders.
The errors are of the same order of magnitude, but the vertex
estimators are much faster and avoid systematic integration errors.

\subsection{Peierls transition in Holstein models}

\begin{figure}[tbp]
\centering
\includegraphics[width=1.0\linewidth]{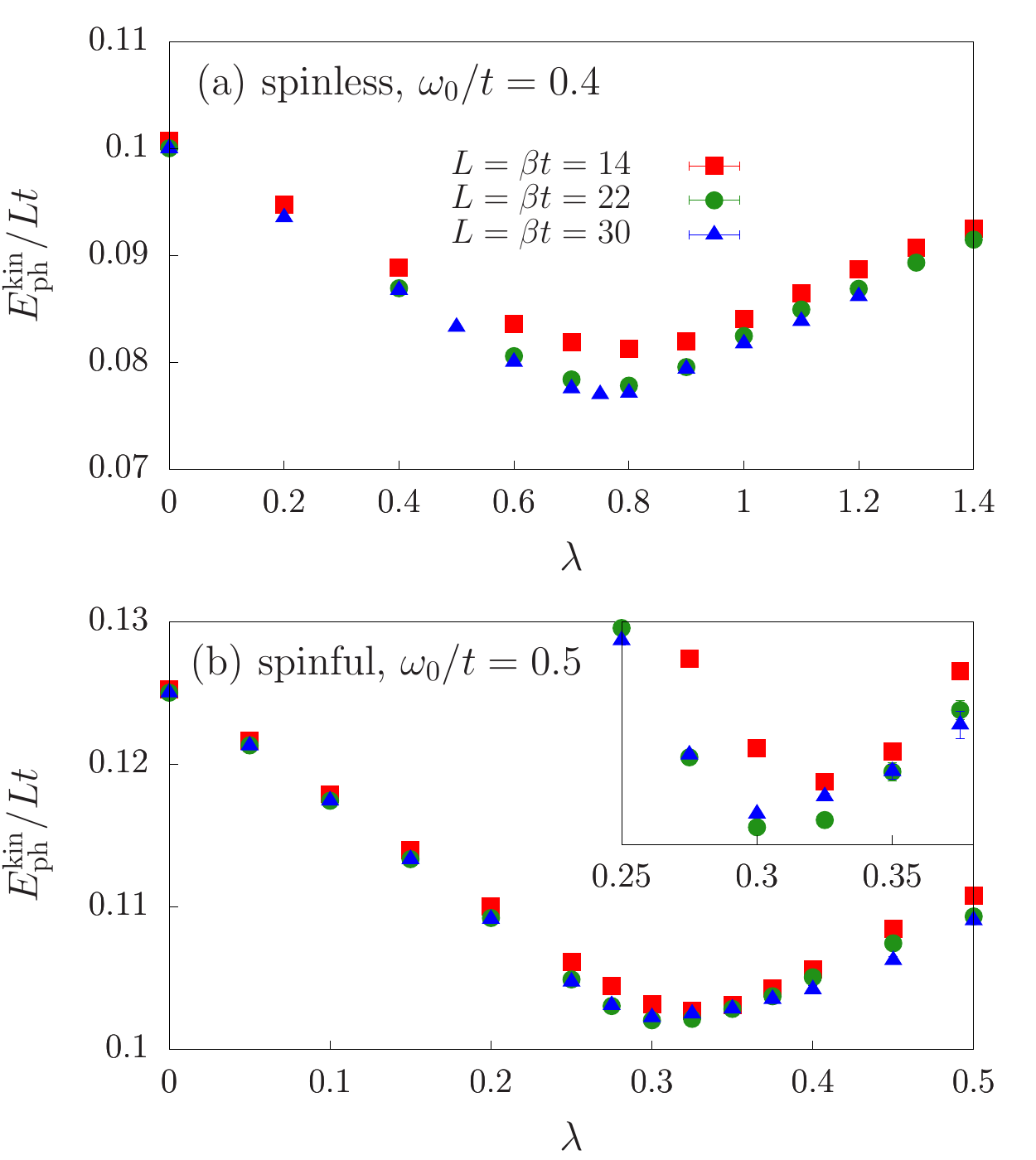}%
\caption{\label{Fig:Energies_adiabatic}
(Color online) Phonon kinetic energy per site
for (a) the spinless Holstein model with $\omz/t=0.4$ and
(b) the spinful Holstein model with $\omz/t=0.5$.
The inset in (b) shows a closeup of the region around the minimum.
}
\end{figure}
\begin{figure*}[tbp]
\centering
\includegraphics[width=\linewidth]{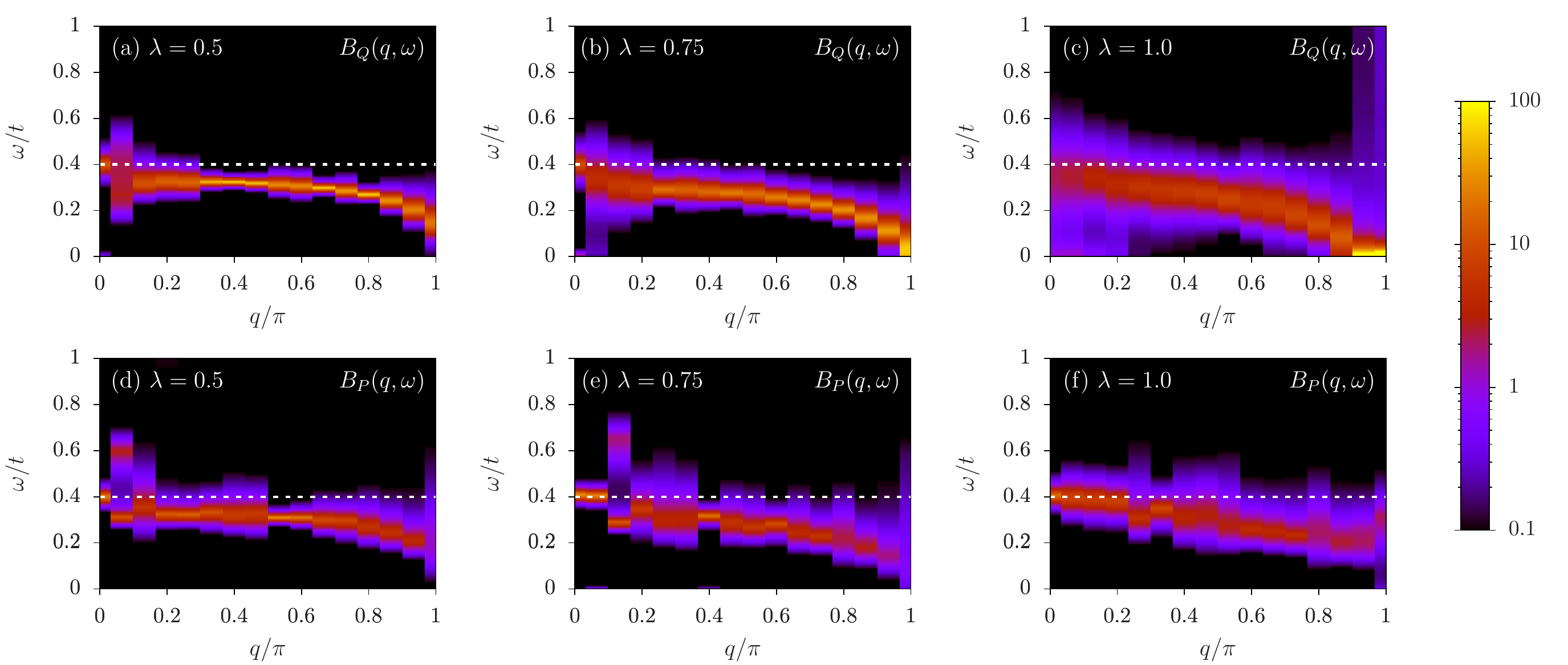}%
\caption{\label{Fig:ph_softening}
(Color online) Phonon spectral functions
$B_Q(q,\omega)$ [(a)--(c)] and $B_P(q,\omega)$ [(d)--(f)] for the spinless
Holstein model.
Dashed lines correspond to $\omz/t=0.4$. Here, $L=\beta
t = 30$. 
}
\end{figure*}

The Peierls quantum phase transition in half-filled spinful Holstein and
Holstein-Hubbard models  has been studied with a number of numerical
techniques (see Ref.~\cite{PhysRevB.92.245132} for a review). While early QMC
results \cite{PhysRevB.27.4302} suggested the absence of a metallic phase,
more recent work has established a phase transition at a nonzero critical
value $\lambda_c$ \cite{PhysRevB.60.7950,PhysRevB.87.075149} in accordance
with functional renormalization group results \cite{Barkim2015}.
However, the exact determination of the phase boundary, as well as the
characterization of the metallic phase in terms of Luttinger liquid
parameters remain open problems \cite{PhysRevB.92.245132}. The difficulties
are associated with the Berezinskii-Kosterlitz-Thouless (BKT) nature of the
transition, so that gaps are exponentially small near $\lambda_c$, and a
small but nonzero spin gap caused by attractive backscattering which is hard
to resolve numerically \cite{PhysRevB.92.245132}. In particular, the spin gap
renders the previously used charge susceptibility \cite{ClHa05} essentially
useless for detecting long-range charge order \cite{PhysRevB.92.245132}.
In contrast, no such complications are encountered for the spinless Holstein
model. Although the quantum phonons still represent a significant numerical
challenge, the phase diagram and the Luttinger liquid parameters have been
determined quite accurately \cite{PhysRevB.73.245120,0295-5075-87-2-27001}. 

Here, we consider alternative diagnostics to detect the Peierls transition, 
namely, the phonon kinetic energy and the fidelity
susceptibility. In addition, we present significantly improved results for
the phonon spectral function over the entire coupling range.

\subsubsection{Phonon kinetic energy}

Figure~\ref{Fig:Energies_adiabatic} shows the
phonon kinetic energy for the spinless Holstein model with
$\omz/t=0.4$ and the spinful Holstein model with $\omz/t=0.5$. 
For both models, $\Ephkintext$ exhibits a distinct minimum as a function of $\lambda$.
In the spinless case, $\Ephkintext$ has
almost converged for the largest system size considered ($L=30$)
and the position of the minimum is consistent
with the previous estimate $\lambda_c \approx 0.7$ \cite{PhysRevB.73.245120}.
While the critical value of the spinful model is still under debate
\cite{PhysRevB.92.245132}, the position of the minimum in
Fig.~\ref{Fig:Energies_adiabatic}(b) suggests a slightly larger value than
in previous results where $\lambda_c\approx0.25$
\cite{0295-5075-84-5-57001,PhysRevB.92.245132}.
The nonmonotonic finite-size dependence of $\Ephkintext(L)$ near $\lambda_c$
in the spinful case is expected to arise from the small but nonzero spin gap
in the metallic phase \cite{PhysRevB.92.245132}.

The minimum in $\Ephkintext$ can be related to the behavior of the 
dynamic charge structure factor $S_{\rho}(q,\omega)$ using the 
sum rules derived in Appendix~\ref{Sec:SumRulesEnergies}. Because of the
density-displacement coupling, $S_{\rho}(q,\omega)$ also contains
contributions from the renormalized phonon dispersion
$\tilde{\omega}(q)$. The minimum of $\Ephkintext$ near $\lambda_c$ arises from the
softening and subsequent hardening of $\tilde{\omega}(q)$ near $q=\pi$
discussed below. Interestingly, a minimum of the phonon kinetic energy is
also observed in the crossover from a large to a small polaron in the Holstein model
\cite{PhysRevB.45.7730,PhysRevB.69.024301}.

The renormalization of $\tilde{\omega}(q)$ was also used in
Ref.~\cite{Creffield2005} to estimate $\lambda_c$
from fits to the phonon Green's function. In our results (see below
and Ref.~\cite{PhysRevB.91.235150})
the value of $\lambda$ at which complete softening of
the phonon mode occurs matches the position of the minimum in $\Ephkintext$.
The latter quantity is easier and faster to calculate with the CT-INT method.
For the spinless Holstein model,
we have also tested this estimator for other phonon frequencies. At
$\omz/t=1$, the position of the minimum in $\Ephkintext$ approaches the
critical coupling $\lambda_c\approx 1.3$ from density-matrix
renormalization group 
calculations \cite{0295-5075-87-2-27001}, but CT-INT simulations become
difficult at these stronger couplings. At $\omz/t=0.1$, we find
considerable finite-size effects even at $\beta t = L =42$ where the
position of the minimum still deviates significantly from
$\lambda_c\approx0.4$ \cite{0295-5075-87-2-27001}.

\subsubsection{Phonon spectral function}

Previous results for the spinless Holstein model suggest that in the
adiabatic regime considered here, the phonon dispersion softens at and around
$q=\pi$ (the ordering wavevector for the Peierls transition) on approaching
$\lambda_c$ from the metallic phase
\cite{PhysRevB.73.245120,SyHuBeWeFe04,Creffield2005,PhysRevB.91.235150,PhysRevB.83.115105}. For
a soft-mode transition, the phonon mode should become completely soft
at $q=\pi$ and $\lambda=\lambda_c$, and subsequently harden
for $\lambda>\lambda_c$. Indications for such a hardening were recently
observed for the spinful Holstein model \cite{PhysRevB.91.235150}, but
a clear identification is complicated by the dominant central peak in the
Peierls phase \cite{PhysRevB.91.235150,PhysRevB.83.115105} and---in the case
of exact diagonalization---the small system sizes accessible at strong
coupling \cite{PhysRevB.73.245120}.

Here, we consider the phonon spectral functions
\begin{align}
B_\alpha(q,\omega)
  =
  \frac{1}{Z}
  \sum_{mn}
  e^{-\beta E_m} 
  \absolute{\bra{m} \hat{O}^{\alpha}_q \ket{n}}^2
  \delta(\omega - \Delta_{nm}) 
\end{align}
calculated either from the displacement [$\alpha=Q$,
Eq.~(\ref{Eq:ph_prop_int_Q})] or the momentum correlation function
[$\alpha=P$, Eq.~(\ref{Eq:ph_prop_int_P})], with $\hat{O}^Q=K^{1/2}\Q{}$,
$\hat{O}^P=M^{-1/2}\P{}$, and $\Delta_{nm}=E_n-E_m$. 

In principle, both spectral functions contain the same information, but
spectral weights may differ significantly. In particular, the Monte Carlo
estimators~(\ref{Eq:ph_prop_Q_vert}) and~(\ref{Eq:ph_prop_P_vert}) may be
subject to different statistical fluctuations that affect the stochastic
analytic continuation \cite{PhysRevB.57.10287,2004cond.mat..3055B}. 

The displacement spectrum  $B_Q(q,\omega)$ in Fig.~\ref{Fig:ph_softening}(a)
reveals the softening of the phonons near $q=\pi$ in the metallic phase.
Near the critical point, the dispersion appears completely soft at $q=\pi$
[Fig.~\ref{Fig:ph_softening}(b)], and the spectrum is dominated by a central
peak at $\omega=0$ associated with the long-range charge order. This peak
grows strongly with $\lambda$ and introduces strong fluctuations 
in the dynamic displacement correlation function~(\ref{Eq:ph_prop_int_Q}) at
all momenta $q$. The fluctuations cause a significant broadening of the
spectrum obtained by analytic continuation, and in particular make it
virtually impossible to resolve finite-frequency contributions at $q=\pi$,
\cf Fig.~\ref{Fig:ph_softening}(c).

To follow the phonon dispersion in the ordered phase, we instead consider 
the spectral function $B_P(q,\omega)$ shown in Figs.~\ref{Fig:ph_softening}(d)--(f).
The use of the momentum correlation function~(\ref{Eq:ph_prop_int_P})
filters out the central mode, and allows us to unambiguously identify the
hardening of the phonon dispersion at $q=\pi$ in the Peierls phase
[Fig.~\ref{Fig:ph_softening}(f)]. Hence, the Peierls transition in the
adiabatic regime can be classified as a soft-mode transition.

\subsubsection{Fidelity susceptibility}

\begin{figure}[tbp]
\centering
\includegraphics[width=\linewidth]{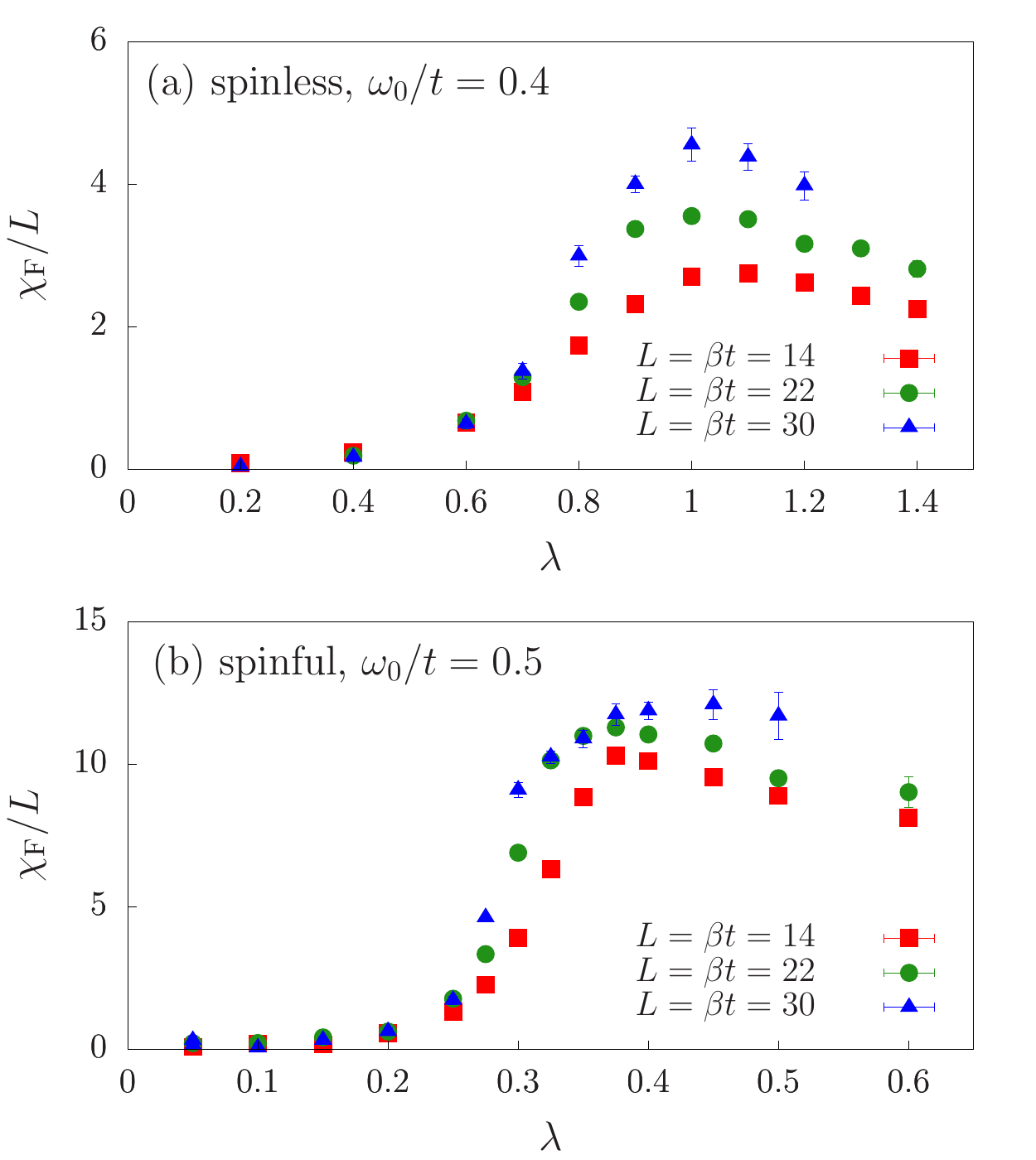}%
\caption{\label{Fig:susfid}
(Color online) Fidelity susceptibility per site for  (a) the spinless
Holstein model with $\omz/t=0.4$ and (b) the
spinful Holstein model with $\omz/t=0.5$. Results were obtained from
Eq.~(\ref{Eq:FS_MC_HS}) with $g^2\to\lambda=g^2/(4Kt)$ and including the
shift discussed after Eq.~(\ref{Eq:FS_MC_HS}).
}
\end{figure}

Using the estimator~(\ref{Eq:FS_MC_HS}) we calculated the fidelity
susceptibility $\chi_\text{F}$ for the spinless and the spinful Holstein
model. The phonon frequencies were chosen as in Fig.~\ref{Fig:Energies_adiabatic}.

Figure~\ref{Fig:susfid}(a) shows $\chi_\text{F}/L$ for the spinless Holstein
model as a function of $\lambda$.  We find a
maximum that grows and shifts to smaller $\lambda$ with increasing $L$. In
contrast, finite-size effects are smaller at weak and strong coupling. In the
thermodynamic limit, a cusp at the critical coupling is expected
for a BKT transition \cite{PhysRevB.91.014418}. For the accessible system sizes, the
position of the maximum deviates significantly from the expected value
$\lambda_c \approx 0.7$ \cite{PhysRevB.73.245120}, in contrast to Fig.~\ref{Fig:Energies_adiabatic}(a). A slow convergence of 
the fidelity susceptibility with system size was previously observed for the
BKT transition in the spin-$\frac{1}{2}$ XXZ chain \cite{PhysRevB.91.014418}.

Results for the spinful Holstein model are shown in Fig.~\ref{Fig:susfid}(b). We again
observe a maximum at intermediate values of $\lambda$ that are significantly
larger than previous estimates $\lambda_c\approx0.25$
\cite{0295-5075-84-5-57001,PhysRevB.92.245132} and the position of the
minimum in Fig.~\ref{Fig:Energies_adiabatic}(b). 
Finite-size effects appear to be less systematic than for the spinless case,
which we attribute to the additional spin gap; the latter is not fully resolved for
small $L$ \cite{PhysRevB.92.245132}. The results in Fig.~\ref{Fig:susfid}(b)
are consistent with a phase transition at a $\lambda_c>0$ and hence a
metallic phase at weak coupling, as reported in previous work. 

\section{Conclusions}
\label{Sec:Conclusions}

The CT-INT quantum Monte Carlo method is particularly useful to simulate
fermion-boson models because the bosons can be integrated out. While
advantageous for simulations, this integration makes it nontrivial to
calculate expectation values of bosonic variables. In this work, we presented
estimators for arbitrary bosonic correlation functions using generating
functionals. As a concrete example, we derived sum rules for the total
energy and the phonon propagator of the Holstein model. Moreover, we
showed that several observables of interest can be measured
directly from the vertex distribution instead of using Wick's
theorem. Additionally, we generalized the QMC estimator
for the fidelity susceptibility \cite{PhysRevX.5.031007} to retarded
boson-mediated interactions, thereby providing a rather general diagnostic to
detect phase transitions.

A comparison of different observables and simulation parameters showed that
statistical errors are of the same order for the vertex estimators and the
estimators based on Wick's theorem. The vertex estimators are easy to
implement, more efficient, and often avoid systematic errors from numerical
integration. These findings complement previous applications in the
context of impurity problems. Our results are general and can be applied to a
variety of other lattice fermion-bosons models. For example, the
possibility to calculate the total energy provides access to the
specific heat. Moreover, the calculation of the charge susceptibility from
the auxiliary Ising spins may be advantageous to detect charge order in
higher dimensions or in Hubbard-type models.

These methodological developments were applied to one-dimensional spinless and spinful Holstein models for
electron-phonon interaction. The phonon kinetic energy was found
to exhibit a minimum related to the renormalization (softening)
of the phonon mode. For intermediate phonon frequencies, the location of the
minimum is consistent with other estimates of the critical point. The phonon
spectral function calculated from the phonon momentum correlator 
reveals the hardening of the phonon mode in the Peierls phase, and thereby
provides evidence for the soft-mode nature of the Peierls transition.
Finally, the fidelity susceptibility exhibits a broad maximum at intermediate
coupling and significant finite-size effects. While it hence does
not provide more accurate critical values in the one-dimensional case
considered, the qualitatively similar behavior observed for the spinless and
the spinful model may be regarded as additional evidence for an extended
metallic phase in the latter.

\begin{acknowledgments}
The authors gratefully acknowledge the computing time granted by the John
von Neumann Institute for Computing (NIC) and provided on the supercomputer
JURECA \cite{Juelich} at Jülich Supercomputing Centre, as well as 
financial support from the DFG Grant Nos.~AS120/10-1 and Ho~4489/4-1
(FOR 1807). We further thank J. Hofmann for helpful discussions.
\end{acknowledgments}

\appendix
\section{Exact relations to the charge spectrum}
\label{Sec:SumRulesEnergies}

For the Holstein model, the phonon propagators
(\ref{Eq:ph_prop_int_Q}) and (\ref{Eq:ph_prop_int_P}) as well as the
energies (\ref{Eq:Epkin})--(\ref{Eq:Eep}) are determined by
the time-displaced charge correlation function
$C_{\rho}(q,\tau-\tau')=\expv{\rhocohan{q}(\tau)\rhocohan{-q}(\tau')}$. The
latter is related to the dynamic charge structure factor
\begin{align}
S_{\rho}(q,\omega)
  &=
  \frac{1}{Z}
  \sum_{mn}
  e^{-\beta (E_m-\mu N_m)} 
  \absolute{\bra{m} \rhoan{q} \ket{n}}^2
  \\\nonumber
  &\hspace*{6em}\times\delta(E_n - E_m - \omega)
\end{align}
via 
$
C_{\rho}(q,\tau)
  =
  \int_0^{\infty} \!\! d\omega \,
  K(\tau,\omega) \, S_{\rho}(q,\omega) $,
where $K(\tau,\omega) = \exp[-\tau\omega] +
\exp[-\left(\beta-\tau\right)\omega]$. Therefore, the entire single-particle
dynamics of the phonons is contained in $S_{\rho}(q,\omega)$. In
particular,
$B(q,\omega)$
is directly related to
$S_{\rho}(q,\omega)$ \cite{PhysRevB.91.235150}.
The energies (\ref{Eq:Epkin})--(\ref{Eq:Eep}) can
be calculated from $S_{\rho}(\omega) = \sum_q S_{\rho}(q,\omega)$ via
{\allowdisplaybreaks
\begin{align}
  \Ephkin
    &=
      \frac{\Ephz}{2}
      - 2 \lambda t
      \! \int_0^{\infty} \!\! d\omega \,
      K_{--}(\omega/\omz, \beta \omz) \, S_{\rho}(\omega) \, ,
	      \label{Eq:Epkin_spec}
	\\
  \Ephpot
    &=
      \frac{\Ephz}{2}
      + 2 \lambda t
      \! \int_0^{\infty} \!\! d\omega \,
      K_{++}(\omega/\omz, \beta \omz) \, S_{\rho}(\omega) \, ,
	      \label{Eq:Eppot_spec}
	\\
  \Eeph
    &=
      - 4 \lambda t
      \! \int_0^{\infty} \!\! d\omega \,
      K_{+}(\omega/\omz, \beta \omz) \, S_{\rho}(\omega) \, ,
	      \label{Eq:Eep_spec}
\end{align}
}
with the kernels ($x=\omega/\omega_0$, $y=\beta\omega_0$, $\omega_0>0$)
\begin{widetext}
\begin{align}
K_{\pm\pm}(x,y)
  =
  \frac{1}{4\pi \left(x^2-1\right)}
  \left\{
  x \tanh(xy/2) \coth(y/2)
  \pm \frac{x y \tanh(xy/2)}{2 \sinh^2(y/2)}
  \mp \frac{2x}{\left(x^2-1\right)}
  \left[ \tanh(xy/2) \coth(y/2) - x^{\mp 1} \right]
  \right\}\,,
\end{align}
\end{widetext}
and $K_+ = K_{++} + K_{--}$, with
\begin{align}
K_+(x,y) 
  =
  \frac{x \tanh(xy/2) \coth(y/2) - 1}{2\pi \left( x^2 -1 \right)}
 \, .
\end{align}

\begin{figure}[htb]
\centering
\includegraphics[width=\linewidth]{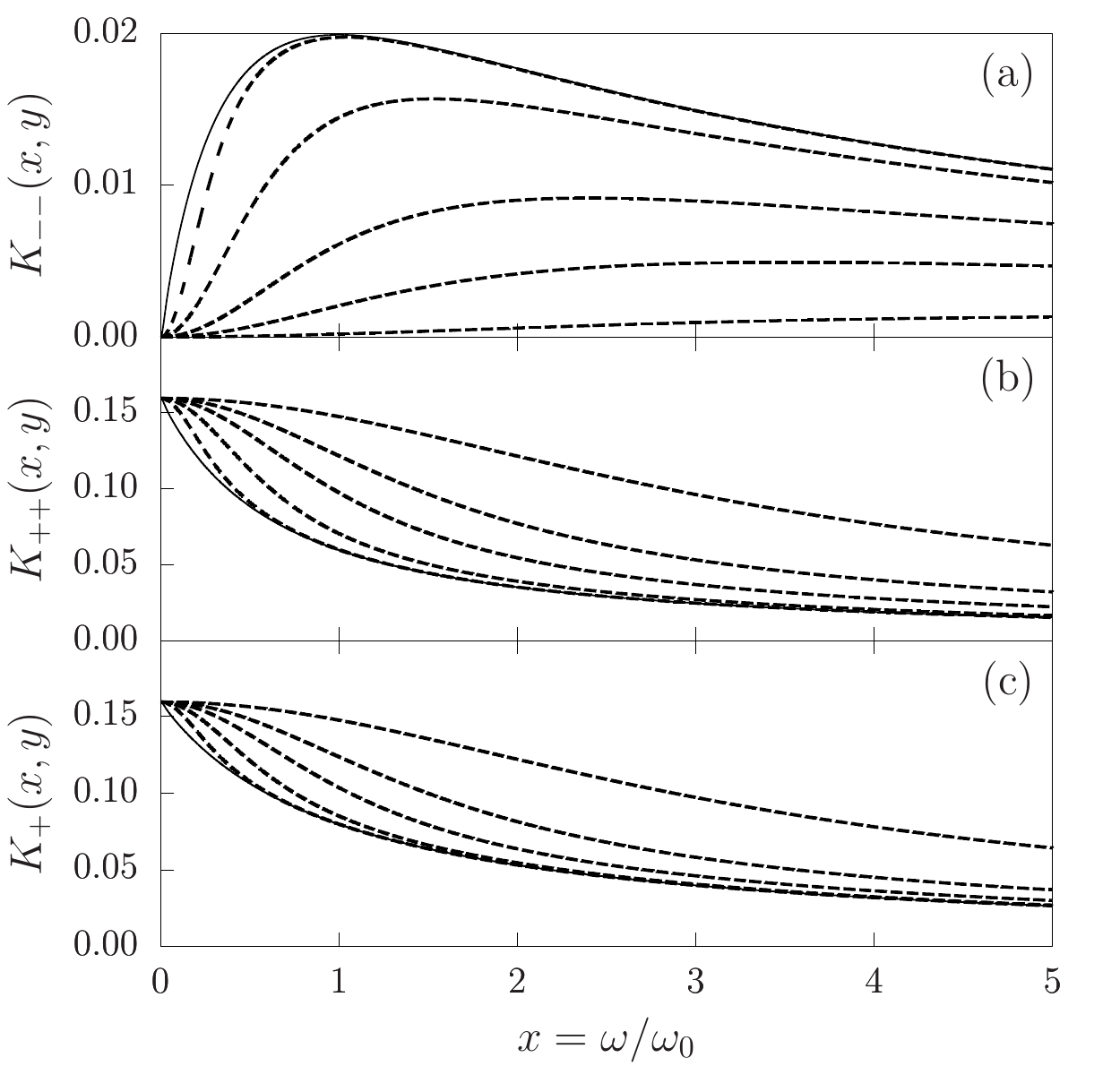}%
\caption{\label{Fig:Kernels}
The kernels $K_{--}$, $K_{++}$, and $K_+$.
Solid lines correspond to $T=0$ results, whereas
dashed lines correspond to $y=\beta\omz=\{10,5,3,2,1\}$ as shown from
the top in (a) and from the bottom in (b)--(c).
}
\end{figure}
The kernels are plotted in Fig.~\ref{Fig:Kernels} for different temperatures.
At $T=0$, $K_{++}$ and $K_{+}$ are largest at $\omega = 0$ and decrease
monotonically with increasing $\omega$, whereas $K_{--}$ is zero at $\omega=0$
and has a maximum at $\omega=\omz$. Therefore, $\Ephpottext$ and $\Eephtext$
mainly capture the charge ordering. In contrast, because $K_{--}$  filters out
the zero-frequency contributions to $S_\rho(\omega)$, $\Ephkintext$ reveals
the softening of the phonons and the opening of the Peierls gap. 
The same reasoning applies to the phonon spectral function.
If calculated from Eq.~(\ref{Eq:ph_prop_int_Q}) it is dominated by the
central mode in the Peierls phase. This mode is filtered out when using Eq.~(\ref{Eq:ph_prop_int_P}).
The kernels
broaden significantly when the temperature becomes comparable to $\omega_0$
but the qualitative behavior for $\omega\ll\omz$
remains unchanged. 

\section{Translational invariance of the vertices}
\label{Sec:TranslVert}

The bosonic estimators from the distribution of vertices can be
substantially improved by exploiting translational invariance in
imaginary time: replacing $\tau_k\to \tau_k+\Delta\tau$ and
$\tau'_k\to \tau'_k+\Delta\tau$ for all vertices $k\in\{1,\dots,n\}$
leaves the weight $W[C_n]$ unchanged. Thereby, we can derive
improved estimators for the bosonic energies (\ref{Eq:Epkin_vert}) and
(\ref{Eq:Eppot_vert}) as well as the phonon propagator
(\ref{Eq:ph_prop_P_vert}).

For the energies (\ref{Eq:Epkin_vert}) and (\ref{Eq:Eppot_vert}), 
translational invariance allows for the transformation
\begin{align}
\label{Eq:prop_repl}
\frac{\Ppm(\tau_k) \Ppm(\tau'_k)}{\Pp(\tau_k - \tau'_k)}
  \, \longrightarrow \,
  \underbrace{
  \frac{1}{\beta} \int_0^{\beta} \!\! d\tau \,
  \frac{
  \Ppm(\tau_k +\tau) \Ppm(\tau'_k +\tau)
  }{
  \Pp(\tau_k - \tau'_k)
  }
  }_{
\Ppmbar(\tau_k - \tau'_k)
  } 
\end{align}
to the averaged propagator ($\tau \in [-\beta,\beta]$)
\begin{align}
\label{Eq:Pbar}
\begin{split}
\Ppmbar(\tau)
=
\frac{1}{2\beta}
&\pm \frac{\omega_0}{4}   \frac{\beta -\absolute{\tau}}{\beta}
\left[\coth(\omega_0\beta/2) - \frac{\Pm(\tau)}{\Pp(\tau)}\right] \\
&\pm \frac{\omega_0}{4}  \frac{\absolute{\tau}}{\beta}
\left[\coth(\omega_0\beta/2) +\frac{\Pm(\tau)}{\Pp(\tau)}\right]
\, .
\end{split}
\end{align}
Since the substitution (\ref{Eq:prop_repl}) applies to time differences of
the same vertex, the computational cost to calculate the energies
remains $\mathcal{O}(n)$. The improvement is particularly noticeable for
$\Ephkintext$ (see Sec.~\ref{Sec:PerformanceTest}).

The simplest way to calculate the phonon propagators (\ref{Eq:ph_prop_Q_vert})
and (\ref{Eq:ph_prop_P_vert}) is to fix the second time argument to $\tau'=0$
and apply Eq.~(\ref{Eq:sumtrick}) to obtain the necessary information from the
vertices in $\mathcal{O}(n N_{\tau})$ operations. Similar to the
equal-time case, especially the estimator for the momentum
correlations can be improved by using translational invariance. However, the rigorous approach of integrating over all
translations increases the computational cost to $\mathcal{O}(n^2
N_{\tau})$ operations since the sums in the first term of
Eq.~(\ref{Eq:sumtrick}) can no longer be calculated independently. This
problem can be overcome by measuring the correlation
functions on an equidistant grid with spacing $\dtauobs$ so that 
translations of all vertices by multiples of $\dtauobs$ are available
and the computational cost remains $\mathcal{O}(n N_{\tau})$. Regardless, translational invariance can
be applied rigorously to the second term in Eq.~(\ref{Eq:sumtrick}). Putting
the contributions of the phonon
propagator together requires another $\mathcal{O}(L^2N_{\tau}^2)$
operations, where an additional factor of $N_{\tau}$ comes from
exploiting translational invariance. This last
step dominates the computational time for vertex measurements (\cf
Sec.~\ref{Sec:PerformanceTest}).


%

\end{document}